\documentclass[physrev,reprint,bibnotes]{revtex4-2}  

\usepackage[T1]{fontenc}
\usepackage[american]{babel}

\usepackage[a4paper,centering,hmargin=1.75cm,vmargin=2cm]{geometry} 

\usepackage{amsmath,amssymb,graphicx,bm,microtype}
\usepackage[dvipsnames]{xcolor}
\usepackage{booktabs}
\usepackage{multirow}
\usepackage[colorlinks,allcolors=blue!50!black]{hyperref}
\usepackage[all]{hypcap}
\usepackage{cleveref}
\usepackage{orcidlink}
\newcommand{\orcid}[1]{\,\orcidlink{#1}}

\usepackage{braket} 
\usepackage[italic,thinc]{esdiff} 
\usepackage[version=4]{mhchem} 
\usepackage{upgreek,textgreek} 
\usepackage{siunitx,textcomp} 
\DeclareSIUnit\bohr{\text {\ensuremath {a}}_{0}}
\DeclareSIUnit\hartree{\text {\ensuremath {E}}_{\mathrm {h}}}

\renewcommand{\thetable}{\arabic{table}} 

\DeclareMathOperator{\Tr}{Tr}

\newcommand{\dg}{\dagger}

\newcommand{\ketbra}[2]{\mathinner{|{#1}\rangle\langle{#2}|}}

\begin{document}

\title{Using bosons to improve resource efficiency of quantum simulation of vibronic molecular dynamics}

\author{Henry L. Nourse\orcid{0000-0003-3610-1105}}
\altaffiliation{These authors contributed equally to this work.}
\affiliation{School of Chemistry, University of Sydney, NSW 2006, Australia}

\author{Vanessa C. Olaya-Agudelo\orcid{0000-0001-5141-0614}}
\altaffiliation{These authors contributed equally to this work.}
\affiliation{School of Chemistry, University of Sydney, NSW 2006, Australia}

\author{Ivan Kassal\orcid{0000-0002-8376-0819}}
\email[Email: ]{ivan.kassal@sydney.edu.au}
\affiliation{School of Chemistry, University of Sydney, NSW 2006, Australia}


\begin{abstract}
Simulating chemical dynamics is computationally challenging, especially for nonadiabatic dynamics, where numerically exact classical simulations scale exponentially with system size, becoming intractable for even small molecules. On quantum computers, chemical dynamics can be simulated efficiently using either universal, qubit-only devices or specialized mixed‑qudit–boson (MQB) simulators, which natively host electronic and vibrational degrees of freedom. Here, we compare the quantum resources required for a qubit-only approach to achieve the same accuracy as an MQB device at simulating nonadiabatic molecular dynamics. We find that MQB simulations require orders-of-magnitude fewer quantum operations than qubit-only simulations, with a one-gate MQB circuit requiring a qubit-equivalent circuit volume of over 400,000 when simulating an isolated molecule, which increases to over ten million when environmental effects are included. These estimates assume perfect qubits and gates, and would increase by additional orders of magnitude if error correction were used for fault tolerance. When errors are small, the advantage of MQB simulators becomes even larger as system size increases. Our results highlight the enormous resource advantages of representing non-qubit chemical degrees of freedom natively, rather than encoding them into qubits.
\end{abstract}

\maketitle

Chemical dynamics is fundamental to understanding the mechanisms of molecular transformations. 
Studying dynamics is particularly important for excited-state processes, where non-equilibrium quantum effects can steer reactions in potentially unexpected ways. Transitions between excited states can lead to nonadiabatic quantum dynamics, which is particularly challenging to understand because it involves simultaneous changes in both electronic and nuclear degrees of freedom, leading to the breakdown of the Born-Oppenheimer approximation that underpins most chemical intuition.
Nevertheless, nonadiabatic dynamics is ubiquitous, playing critical roles in processes such as photosynthesis~\cite{gatti2014molecular}, atmospheric chemistry~\cite{curchod2024perspective}, and solar energy conversion~\cite{ponseca2017ultrafast}.

Accurately simulating nonadiabatic dynamics is challenging, even for small molecules, because the computational cost grows exponentially with system size for numerically exact, fully quantum algorithms. The cost arises from the need to accurately treat multiple coupled potential-energy surfaces~\cite{cederbaum1981multimode,worth2004beyond,domcke2004conical}. Therefore, accurate state-of-the-art simulations are limited to chemical systems with a few tens of nuclear degrees of freedom~\cite{aworth2008using,curchod2018ab,nelson2020non-adiabatic,cigrang2025roadmap}, and the situation is even more severe in the condensed phase, where one must model the influence of a large environment through open-system dynamics~\cite{bonfanti2012compact,vanhaeften2023propagating,curchod2018ab,crespo-otero2018recent,nelson2020non-adiabatic}.

Quantum computers offer efficient simulation alternatives. Qubit-only simulators have universal applicability, since any quantum system can be simulated efficiently on a quantum computer~\cite{lloyd1996universal,cao2019quantum}. However, applying them to molecular dynamics involves discretizing continuous variables into qubits~\cite{kassal2008polynomial-time,Sawaya2020,ollitrault2020nonadiabatic,ollitrault2021molecular,jornada2025comprehensive,motlagh2024quantum}, which introduces substantial overhead in both the number of qubits and the circuit size. Nevertheless, qubit-only platforms carry the long-term prospect of fault-tolerant implementation because error-corrected qubits can execute deep circuits with controlled accuracy.

By contrast, mixed-qudit-boson (MQB) simulators naturally describe vibronic dynamics by mapping electronic states to qudits and nuclear motion to bosonic modes, such as the vibrations of trapped ions~\cite{MacDonell2021}. They have performed analog simulations of nonadiabatic dynamics in a range of chemical systems~\cite{macdonell2023predicting,valahu2023direct,navickas2025experimental,whitlow2023quantum}, including dissipative processes of molecules interacting with an environment~\cite{sun2025quantum,so2024trapped-ion,navickas2025experimental}, which would incur significant overhead in qubit-only simulators~\cite{digitalOQS2011Wang, OQSalgo2015,OQSalgo2016,childs2017efficient,cleve2017efficient,digitalOQS2021,li2023simulating,suri2023twounitary,LinLin2024,pocrnic2025quantum}. MQB devices promise to simulate nonadiabatic dynamics with high resource efficiency, with the potential of scaling to classically intractable systems with near-term technology. However, because they lack error correction they are susceptible to noise and the accumulation of errors.

Comparing the computational cost of MQB simulators and qubit-only computers is challenging because the two paradigms differ in how they represent quantum states, handle continuous variables, and scale with system size. They also have different sources of error, with MQB simulators suffering uncorrectable noise and qubit-only algorithms having greater algorithmic overhead and approximation error when bosonic degrees of freedom are discretized into qubits. Previous analyses of MQB simulation, primarily in lattice-gauge-theory and fermion-boson settings, have shown that access to native bosonic operations can reduce the number of quantum operations because compiling them onto qubit-only devices introduces long gate sequences~\cite{davoudi2021toward,crane2024hybrid,kumar2025digitalanalog,chiari2025abinitio}. However, these approaches do not resolve the key issue for MQB simulators, i.e., determing when the accumulated hardware noise remains low enough that their resource advantage over qubit-only implementations survives. Answering this question requires a fair comparison of MQB and qubit-only approaches carrying out simulations with equal accuracy.

Here, we develop a systematic framework for comparing quantum resource costs between the two quantum-simulation approaches in the context of molecular dynamics. We quantify the quantum resources required in both approaches for simulating chemical dynamics to the same error. To ensure fairness, the comparison is made using the most resource-efficient method for each simulator, yielding conservative benchmarks. Using pyrazine as a case study, we find that MQB simulation requires dramatically fewer resources than qubit-only simulations to achieve the same error and that this advantage increases with system size, as long as errors are small. This is the case both for isolated molecules and, even more so, for molecules interacting with an environment. 

\section{Comparing quantum resources}

Our approach to comparing quantum resource requirements on different types of simulators is as follows: (1)~determining the resources to be counted for each simulator, (2)~determining the error sources in both, and (3)~counting the resources required for each simulator to achieve the same error in the desired observable.

The resource cost for quantum simulation has two broad components: computational memory and computational time. Memory is the number of information-carrying degrees of freedom (qubits, qudits, or bosonic modes) required to encode the relevant dynamics. It depends on the size of the molecular system, how it is mapped onto the simulator, and any memory overhead of the simulation algorithm. Time is the number of quantum operations (gates) that must be performed to evolve an initial state into the final state. This cost depends on the size of the system, the simulation algorithm, and the desired error in the final state.

To compare resource efficiency between quantum simulators, we employ a quantum volume metric that unifies memory and time costs~\cite{moll2018quantum,cross2019validating,blume-kohout2020volumetric,baldwin2022re-examining}. We define the volume as the product of the memory and the time costs above. To compare resource efficiencies, for a given MQB volume (MQBV), we define the qubit-equivalent circuit volume (QECV) as the volume required for a qubit-only device to reproduce the same simulation task with the same error. The relative resource savings achieved by the MQB approach are then quantified by the advantage 
\begin{equation}
    A = \frac{\textrm{QECV}}{\textrm{MQBV}}.
\end{equation}

To ensure a comparison of comparable simulation outcomes in the two approaches, simulation costs must be compared at fixed accuracy. The accuracy can be measured in a variety of ways~\cite{Nielsen}, and the choice can affect the relative cost of the two simulations. In examples below, we track the error in the fidelity of the final state and in a chemically relevant observable, but our approach can be applied to any error metric.

There are two broad classes of errors in both MQB and qubit-only simulation. The first are experimental errors, being inaccuracies in gate execution and decoherence induced by the environment. The second are algorithmic errors, caused by mathematical approximations in the simulation algorithms. In principle, both would be accounted for in both types of simulations.

However, to simplify the discussion and highlight the key differences between MQB and qubit-only algorithms, we assume that the only source of error in MQB simulation is experimental error and in qubit-only simulation algorithmic error.
The motivation for this distinction is that MQB simulation lacks error correction, so it accumulates error from experimental inaccuracies. Over time, this error will exceed any controllable algorithmic errors.
In contrast, qubit-only simulations can, in principle, use error-correcting codes, in which multiple physical qubits encode higher-level logical qubits that suffer lower error rates~\cite{lidar2013quantum,terhal2015quantum,girvin2023introduction,Katabarwa2024,ErrorCorrectionZoo}. Gates on logical qubits then need to be decomposed into operations on the physical qubits, which incurs resource overheads. 
Therefore, the conservative approach to comparing the two errors is to compare the experimental error of MQB simulation with qubit-only algorithmic error. This implies that our qubit-only estimates are for perfect qubits and gates, free of experimental error. If the low experimental error is achieved through error correction, that overhead would increase the qubit-only resources.

\section{Vibronic molecular dynamics}\label{sec:VC_dynamics}

Vibronic molecular dynamics can be described by vibronic coupling (VC) Hamiltonians, which account for the coupling between electronic and vibrational degrees of freedom. VC Hamiltonians are Taylor expansions of the form~\cite{worth2004beyond,domcke2004conical} (we set $\hbar = 1$ throughout)
\begin{align} \label{eq:ham-vc}
	H_{\mathrm{VC}} &= \frac{1}{2} \sum_{j=1}^M \omega_j \left( P_j^2 + Q_j^2 \right) + \sum_{n,m=1}^D W_{n,m} \ketbra{n}{m}, \\
	W_{n,m} &= c_0^{(n,m)} + \sum_j c_j^{(n,m)} Q_j + \sum_{j,k} c_{j,k}^{(n,m)} Q_j Q_k + \ldots ,
\end{align}
where $\ket{n}$ is the $n$th diabatic electronic state; $P_j$ and $Q_j$ are the dimensionless momentum and position of the $j$th vibrational mode with frequency $\omega_j$; $D$ is the number of electronic states; $M$ is the number of vibrational modes; and $c_0^{(n,m)}$, $c_j^{(n,m)}$, and $c_{j,k}^{(n,m)}$ are the expansion coefficients of the molecular potential energy about the reference nuclear geometry (usually the minimum of the ground electronic surface).

Information about the system's dynamics is obtained by measuring observables, such as electronic-state populations, vibrational distributions, and spectroscopic correlation functions~\cite{schatz2002quantum}. To determine these quantities for an isolated molecule, the initial molecular state $\rho_0$ is propagated in time under the unitary operator
\begin{equation} \label{eq:unitary-vc}
	U(t) = \exp(-iH_{\mathrm{VC}}t),
\end{equation}
to the final state $\rho(t) = U(t)\rho_0U^{\dagger}(t)$, on which any observable can be measured. For concreteness, we focus on diabatic population dynamics
\begin{equation}
	P_n(t) = \Tr(\rho(t) \ket{n}\bra{n} ),
\end{equation}
which are central to processes such as photoisomerization, exciton transfer, and internal conversion.

Molecular dynamics rarely occurs in isolation, instead taking place under continuous interaction with an environment. Even weak system-environment couplings can substantially modify dynamics, leading to phenomena such as decoherence, nonadiabatic transitions, and energy and charge transfer~\cite{nitzan2006chemical}. Many theoretical approaches have been developed to model these effects, but we restrict ourselves to Markovian dissipation described by the Lindblad master equation~\cite{nitzan2006chemical},
\begin{equation} \label{eq:lindblad-vc}
	\diff{{\rho}}{t} = -i [ {H}_\mathrm{VC},{\rho} ] + \sum_{i=1}^J  \gamma_i \mathcal{D}\left[{V}_i\right]\rho,  
\end{equation}
where $\gamma_i$ is the rate at which the dissipator $\mathcal{D}\left[V_i\right]$ acts on the density matrix as
\begin{equation}
	\mathcal{D}\left[V_i\right]\rho = {V}_i {\rho} {V}_i^{\dag} - \tfrac{1}{2}\{{V}_i^{\dag}{V}_i,{\rho}\}.
	\label{eq:LindbladRates}
\end{equation}

In chemical systems, the Lindblad equation usually arises from the assumption of weak and Markovian system-environment coupling, together with the secular approximation~\cite{nitzan2006chemical}. In that limit, the dominant dissipative processes are population transfers and dephasing, for both electronic and vibrational degrees of freedom. In particular, they are: radiative electronic relaxation described by $\mathcal{D}[\ket{n}\bra{m}]$; vibrational heating and cooling in mode $j$, respectively described by $\mathcal{D}[a_j^{\dagger}]$ and $\mathcal{D}[a_j]$, where the annihilation operator for mode $j$ is $a_j = (Q_j + iP_j)/\sqrt{2}$; electronic pure dephasing described by $\mathcal{D}[\ket{n}\bra{n}]$; and vibrational pure dephasing in mode $j$ described by $\mathcal{D}[a_j^{\dagger} a_j]$.

\section{MQB simulation} \label{sec:mqb-simulation} 

MQB simulators~\cite{MacDonell2021} use both qudit and bosonic degrees of freedom, which can be implemented on a range of hardware, including ion traps~\cite{mlmer1999multiparticle,leibfried2003experimental,macdonell2023predicting,valahu2023direct,whitlow2023quantum,so2024trapped-ion,kang2024seeking,sun2025quantum}, circuit quantum electrodynamics~\cite{blais2004cavity,mischuck2013qudit,krastanov2015universal,liu2021constructing,eickbusch2022fast,wang2023observation}, and neutral atoms~\cite{schlosser2001sub-poissonian,sortais2007diffraction-limited,kaufman2012cooling,lshaw2025erasure}. MQB devices natively encode the molecular degrees of freedom in $H_{\mathrm{VC}}$, with qudits representing the electronic states and bosons the vibrational modes. Hence, $H_{\mathrm{VC}}$ can be directly mapped onto an MQB simulator, with $H_{\mathrm{MQB}} = F H_{\mathrm{VC}}$, where $F$ is a scaling factor that accounts for differences in energy scales. An MQB simulation consists of preparing the simulator in an initial state, allowing it to evolve under $H_\mathrm{MQB}$ for a desired time, and measuring the observable of interest.

An MQB simulator can implement $H_{\mathrm{VC}}$ using the Hamiltonian~\cite{MacDonell2021} 
\begin{align}
	H_\mathrm{MQB} & = \frac{1}{2}\sum_{j=1}^{M} \delta_j a^\dg_j a_j + \sum_{n=1}^{D} \chi_n \ket{n}\bra{n} \nonumber \\
	& \quad + \sum_{ n\neq m} \Omega_{n,m} \ket{n}\bra{m} \nonumber \\
    & \quad +  \sum_n \sum_j \Theta_{n,j}' (a^\dg_j + a_j) \ket{n}\bra{n} \nonumber \\
	& \quad +  \sum_{n, m} \sum_{j} \Omega'_{n,m,j} (a^\dg_j + a_j) \ket{n}\bra{m} + \ldots,
	\label{eq:closed-system-simulatorH}
\end{align}
where, with some abuse of notation, we use $a_j$ to also denote the annihilation operator of the $j$th bosonic mode and $\ket{n}$ is a qudit state. All parameters can be programmed to reproduce relevant VC parameters and simulate a molecule of interest~\cite{MacDonell2021}. For example, for an ion-trap simulator, the parameters are programmed by adjusting the appropriate laser quantities~\cite{MacDonell2021}: $\delta_j$ arises from a laser's detuning from resonance with mode $j$, $\chi_n$ is a time-independent AC Stark shift, $\Omega_{n,m}$ can be controlled by changing the laser intensity or its detuning from an auxiliary electronic transition, $\Theta_{n,j}'$ corresponds to a time-dependent AC Stark shift from a pair of non-copropagating lasers, and $\Omega_{n,m,j}'$ is proportional to the Rabi frequency of a M{\o}lmer-S{\o}rensen interaction. 

MQB simulators are resource efficient for vibrational molecular dynamics in both memory and time. The memory cost comes from two types of degrees of freedom: the number of qudits is always one and the number of bosons equals the number of vibrational modes in $H_{\mathrm{VC}}$. Hence, the memory cost scales linearly with the number of molecular vibrational modes~\cite{MacDonell2021}. Depending on the underlying hardware, auxiliary qudit(s) and boson(s) might be used to implement certain gates~\cite{eickbusch2022fast,liu2025hybrid}. The time cost comes from single-qudit rotations, boson operations, and qudit-boson entangling operations, such as bosonic conditional displacements. Many of these operations can be combined into compound control pulses, which we count as one gate. Not all terms in $H_{\mathrm{VC}}$ are found in all types of MQB hardware, but they can be synthesized~\cite{MacDonell2021,park2018deterministic,sutherland2021universal,park2024efficient,albornoz2024oscillatory,liu2025hybrid,chalermpusitarak2025programmable} from native gates, which incurs an approximation error.

Extending the MQB approach to simulate molecules in an environment has the same resource cost as MQB simulation of unitary dynamics, but can be more accurate because environmental noise can be used as a resource~\cite{MacDonell2021,navickas2025experimental,Olaya_2025}. Our goal is to simulate molecular dynamics under the master equation \cref{eq:lindblad-vc}, which MQB simulators can implement naturally~\cite{MacDonell2021,navickas2025experimental,Olaya_2025} using two sources of dissipation: the intrinsic dissipation of the simulator itself and injected controllable dissipation, which can be engineered.

Native dissipation in the simulator can be classified as either \textit{usable} or \textit{unusable}~\cite{Olaya_2025}, depending on whether it can be harnessed to simulate dissipative molecular dynamics. For molecules interacting with an environment as described by \cref{eq:lindblad-vc}, usable dissipation is the Markovian simulator noise characterized by Lindblad dissipators $\mathcal{D}[V_i]$ and occurring at a native rate $\gamma_i^\mathrm{nat}$, which can be measured in preliminary experiments. The unusable dissipation describes simulator dissipation that does not correspond to molecular dissipation processes.

As in simulations of isolated molecules, molecular dissipation rates are scaled to the MQB device as $\gamma_i^\mathrm{MQB} = F \gamma_i$. However, when native dissipation is insufficient to reproduce the desired molecular dissipation, additional controllable dissipation of the same type $\mathcal{D}[V_i]$ can be injected, with rate $\gamma_i^{\mathrm{inj}}$. The total rate is then
\begin{equation}\label{eq:inj}
\gamma_i^{\mathrm{MQB}} = \gamma_i^{\mathrm{nat}} + \gamma_i^{\mathrm{inj}} = F \gamma_i.
\end{equation}
There are protocols to implement molecular dissipative processes using controllable noise injection. For example, in trapped-ion architectures~\cite{navickas2025experimental,Olaya_2025}, radiative electronic relaxation, $\mathcal{D}[\ket{n}\bra{m}]$, can be engineered using optical pumping to artificially shorten the lifetime of an excited state. Vibrational heating, $\mathcal{D}[a_j^{\dagger}]$, and cooling, $\mathcal{D}[a_j]$, can be implemented via resolved-sideband laser interactions applied to an ancillary ion that shares vibrations with the MQB simulator ions. Electronic, $\mathcal{D}[\ket{n}\bra{n}]$, and vibrational, $\mathcal{D}[a_j^{\dagger} a_j]$, pure dephasing can be injected using stochastic fluctuations of the energies of the electronic and vibrational states.

There are several ways to choose the scaling factor~$F$, depending on the dissipators involved and the fraction of injected noise included in the simulation~\cite{Olaya_2025}. Here, we only include one type of usable dissipation in the simulation, with rate $\gamma_i^{\textrm{MQB}}$. To maximize the simulation duration, we choose $F=\gamma_i^{\mathrm{nat}}/\gamma_i$ so that this native dissipation is fully used, thereby eliminating the need for injected dissipation, $\gamma_i^{\mathrm{inj}} = 0$~\cite{Olaya_2025}. To distinguish the native dissipation used as a resource from unwanted dissipation that propagates into experimental errors, we denote the latter by $\gamma_i^{\mathrm{err}}$.

\section{Qubit-only simulation} \label{sec:qubit-simulation}

Qubit-only quantum simulators encode both the electronic states and the vibrational modes of $H_{\mathrm{VC}}$ into qubits. All encodings of the vibrational modes are approximate because they represent an infinite-dimensional harmonic-oscillator Hilbert space in a finite number of qubits; however, the corresponding error can be systematically reduced at the cost of more quantum resources.

Different encodings---e.g., unary, binary, and Gray, whether in real space or Fock space---give different qubit and gate costs, sometimes offering tradeoffs between the two~\cite{Sawaya2020}. Binary and Gray encodings minimize the number of qubits but typically require more gates, while unary uses more qubits to reduce gate counts. Real-space representations give simple position operators at the cost of complicated kinetic terms, whereas Fock-space representations can be more efficient for small oscillator displacements, where only low Fock states are occupied.

The memory cost of our qubit-only simulations are the numbers of qubits needed to encode the electronic and vibrational modes of $H_{\mathrm{VC}}$. We use a Fock-space representation with the Gray encoding because it uses the fewest qubits and is more resource efficient than binary for position operators $Q_j$~\cite{Sawaya2020}. We assign a register of qubits to the electronic states (labeled $x_0$) and a register to each vibrational mode ($x_1,\ldots,x_M$), which have been truncated to a $d_j$-level Fock space. The electronic states require $s_0 = \lceil \log_2 D \rceil$ qubits and each vibrational mode $s_j = \lceil \log_2 d_j \rceil$ qubits.

With that mapping, $H_{\mathrm{VC}}$ can be mapped onto qubits by decomposing each of its terms into a sum of Pauli strings. On a single qubit, the Pauli operators $\{I, \sigma_x, \sigma_y, \sigma_z \}$ form a basis for linear operators, and on $s_j$ qubits, the basis is formed by the $4^{s_j}$ tensor products of these. Hence, any operator $A_i$ acting on register $i$ can be Gray-encoded as the Pauli sum~\cite{Sawaya2020}
\begin{equation} \label{eq:qubit-operator-map}
	\mathcal{G}[A_i] = \sum_{k=1}^{4^{s_i}} \alpha_{i,k} \mathcal{P}_{i,k},
\end{equation}
where the Pauli string $\mathcal{P}_{i,k} = \bigotimes_{\ell \in x_i} \sigma_{k,\ell}^{(i)}$ is a product of Pauli operators $\sigma_{k,\ell}^{(i)}$ acting on qubit $\ell$ of register $x_i$, and $\alpha_{i,k} = \Tr(P_{i,k} A_i) / 2^{s_i}$. In practice, $\mathcal{G}[A_i]$ can have drastically fewer Pauli strings than $4^{s_i}$ if it is, e.g., diagonal, local, or sparse.

To map $H_{\mathrm{VC}}$ onto qubits, $\mathcal{G}[\cdot]$ is applied to each term in $H_{\mathrm{VC}}$, except that we combine $P_j^2$ and $Q_j^2$ into $n_j = (P_j^2 + Q_j^2) / 2$ because diagonal operators require fewer gates. For each term, we decompose the operators acting on different degrees of freedom separately and multiply the results together, e.g., $\mathcal{G}[\ket{n}\bra{m} Q_i^r Q_j^v] = \mathcal{G}[\ket{n}\bra{m}] \mathcal{G}[Q_i^r] \mathcal{G}[Q_j^v]$. After grouping equivalent Pauli strings and removing those proportional to the identity (which do not affect the dynamics), $H_{\mathrm{VC}}$ becomes
\begin{equation} \label{eq:ham-vc-qubit}
	H_{\mathrm{qubits}} = \sum_{k=1}^L \alpha_k \mathcal{P}_k.
\end{equation}
where $\mathcal{P}_k$ is a Pauli string with coefficient $\alpha_k$ and $L$ is the total number of Pauli strings.

\subsection{Isolated molecule}

To simulate the unitary dynamics under $H_{\mathrm{qubits}}$, various quantum-simulation algorithms could be used, including Trotter-based algorithms~\cite{lloyd1996universal, kassal2008polynomial-time} or qubitization-based approaches~\cite{Low2019,gilyen2019quantum,martyn2021grand}. These exhibit different time and space costs, as well as different asymptotic scalings, and the choice of the best algorithm depends on the system under study and the available resources. 

Here, we adopt Trotter-based algorithms because they tend to have lower overheads for small systems compared to the asymptotically faster qubitization-based approaches~\cite{childs2021theory,childs2019nearly,bosse2025efficient}. In Trotterization, the total simulation time $T$ is discretized into $N$ steps of length $\Delta t = T/N$. The unitary evolution up to time $T$ is then $U(T) \approx (U^{(p)}(\Delta t))^N$, where $U^{(p)}(\Delta t)$ is the $p$th order Trotter approximation for each time step. At first order,
\begin{equation} \label{eq:trotter-first-order}
	U^{(1)} (\Delta t) = \prod_{k=1}^L U_k (\Delta t),
\end{equation} 
where $U_k (\Delta t) = \exp(-i\, \alpha_k \mathcal{P}_k\, \Delta t)$. Higher-order Suzuki-Trotter decompositions~\cite{suzuki1990fractal,suzuki1991general} have better asymptotic scaling but can require more quantum resources at smaller errors~\cite{childs2018first,childs2021theory,morales2025selection}.
To fairly compare with MQB simulation, the Trotter order that has the lowest resource count should be used. We use first-order because it is simple to analyze and performs comparably to second-order at errors achieved by MQB simulators (see \cref{appen:higher_trotter}). 
Another Trotter-based quantum algorithm for VC models~\cite{motlagh2024quantum} achieves comparable costs to our method using a different bosonic representation and quantum circuit.

With this choice of algorithm, we can estimate the required qubit-only time cost. Each $U_k(\Delta t)$ can be decomposed into a CNOT ladder~\cite{Nielsen,cowtan2020phase,wetering2021constructing,elisa2025measurement-based} containing one $R_z\!\left(2 \alpha_k \Delta t\right)$ gate and $2(p_k-1)$ CNOT gates, where $p_k$ is the number of non-identity single-qubit Pauli gates in the Pauli string $\mathcal{P}_k$~\cite{Sawaya2020}. Therefore, in a single Trotter step, the total number of $R_z$ gates is $L$ and the total number of CNOT gates is $\sum_{k=1}^L 2(p_k-1)$.  However, it is thought that the number of CNOT gates can be reduced through optimization by up to a factor of~3~\cite{Sawaya2020}; therefore, to be conservative in our comparison, we report the CNOT cost as $\sum_{k=1}^L 2(p_k-1)/3$. For simulation to the final time $T$, the total number of gates is $N$ times the number of gates for a single step.

\begin{figure}
    \centering
    \includegraphics[width=\columnwidth]{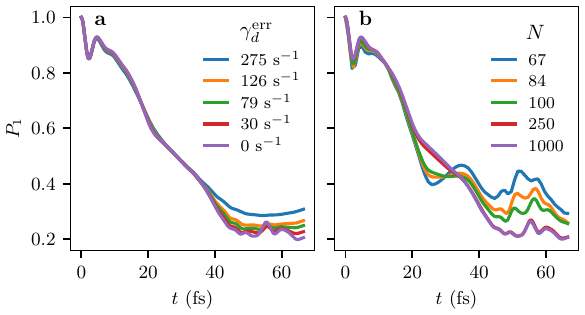}
    \caption{Simulated population dynamics of the pyrazine bright state $\pi\pi^*$ subject to different sources of error, using an LVC model with two electronic states and two vibrational modes.
    \textbf{(a)}~Dynamics in an MQB simulator, subject to pure vibrational dephasing with rate $\gamma_d^{\mathrm{err}}$ on both vibrational modes. 
    \textbf{(b)}~Dynamics on a qubit-only simulator, subject to first-order Trotter error due to discretization to $N$ steps.
    }
    \label{fig:dynamics}
\end{figure}

\subsection{Molecule in environment}

Qubit-only devices can also simulate open quantum systems, often by using additional qubits to simulate the environment~\cite{digitalOQS2011Wang, OQSalgo2015, OQSalgo2016, childs2017efficient, cleve2017efficient, digitalOQS2021, li2023simulating, suri2023twounitary, LinLin2024, pocrnic2025quantum}. Here, we use an algorithm that unravels the Lindblad dynamics into a stochastic differential equation (SDE) that can be solved using unitary dynamics in an enlarged Hilbert space~\cite{LinLin2024}. It has low overhead and allows the use of the same Trotter-based simulation approach we use for an isolated molecule. Specifically, we use the dilated Hamiltonian
\begin{equation} \label{eq:ham-dilated-first-order}
	H_{\mathrm{dil}} = \begin{pmatrix}
		\sqrt{\Delta t} H_{\mathrm{qubits}} & V_1 & \ldots & V_J \\
		V_1^{\dagger} & 0 & \ldots & 0 \\
		\vdots & \vdots & \ddots &  \vdots \\
		V_J^{\dagger} & 0 & \ldots & 0
	\end{pmatrix},
\end{equation}
where $V_1, \ldots, V_J$ are the jump operators from \cref{eq:lindblad-vc}. $H_{\mathrm{dil}}$ is a block matrix whose entries are $\sqrt{\Delta t}H_{\mathrm{qubits}}$ and $V_i$, and a register of $\lceil \log_2(J+1) \rceil$ ancilla qubits (labeled $b$) is introduced for a binary encoding of the block index.

A discrete-time approximation to the evolution of $\rho$ under \cref{eq:lindblad-vc} is~\cite{LinLin2024}
\begin{equation} \label{eq:lindblad-density-update}
	\rho_{m+1} \approx \Tr_b \left( e^{-i\sqrt{\Delta t} H_{\mathrm{dil}}} \,\ket{0_b}\bra{0_b} \otimes \rho_m \,e^{i\sqrt{\Delta t} H_{\mathrm{dil}}} \right),
\end{equation}
where $\rho_m$ is the reduced density matrix of the molecule at time $t = m \Delta t$ and $\ket{0_b}$ are the ancilla qubits that are measured, discarded, and reset after each time step. The required time evolution generated by $H_{\mathrm{dil}}$ can be achieved with the same method used to time-evolve $H_{\mathrm{qubits}}$. Hence, we time evolve $H_{\mathrm{dil}}$ using a first-order Trotter algorithm, as in \cref{eq:trotter-first-order}. 

Counting resources for the time evolution under $H_{\mathrm{dil}}$ is similar to counting resources for evolution under $H_{\mathrm{qubits}}$. The memory cost is the same as for the simulation of the isolated molecule, plus an additional $\lceil \log_2(J+1) \rceil$ ancilla qubits for the bath register. The time cost is counted in the same way as for \cref{eq:trotter-first-order}, except that the unitary dynamics is now governed by $H_{\mathrm{dil}}$, so the number of $R_z$ and CNOT gates will be higher. This increase occurs because of the increased dimension of $H_{\mathrm{dil}}$ (which requires more Pauli strings and thus more $R_z$ gates) and because the CNOT ladder includes additional entangling operations with the bath register.

In our simulations of $H_{\mathrm{qubits}}$ and $H_{\mathrm{dil}}$, we assume perfect qubits and gates. Hence, there are two classes of errors in our qubit-only simulations: the error from truncating the vibrational modes to $d_j$-level Fock spaces and the Trotter error in describing the dynamics under $H_{\mathrm{qubits}}$ or $H_{\mathrm{dil}}$. Because we are interested in quantifying the necessary quantum resources due to algorithmic error, in our simulations we choose $d_j=32$, which is large enough that the truncation error is negligible compared to the Trotter error. The algorithmic error can be systematically improved by increasing the Trotter number $N$, i.e., decreasing the time step $\Delta t$. 

\begin{figure}
    \centering
    \includegraphics[width=\columnwidth]{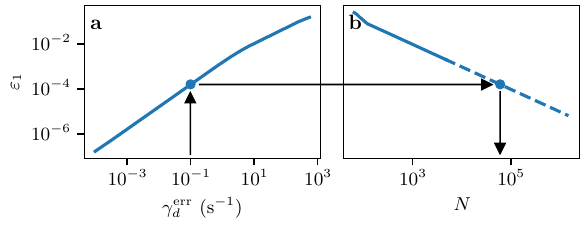}
    \caption{Matching errors between MQB and qubit-only simulations of isolated pyrazine. 
    \textbf{(a)} MQB error of the excited-state population $P_1$ as a function of the vibrational pure-dephasing rate $\gamma_d^{\mathrm{err}}$. 
    \textbf{(b)} Same error for a qubit-only simulator, as a function of the Trotter number $N$. Arrows depict an example of error matching between MQB and qubit-only simulations: for $\gamma_d^{\mathrm{err}} = \SI{0.1}{\per\second}$, the MQB simulation error of $\varepsilon_1 = \num{1.6e-4}$ is matched to the same qubit-only error, which occurs at $N=\num{61000}$. Dashed line is extrapolation.
    }
    \label{fig:error_matching}
\end{figure}

\section{Example: Pyrazine}

To demonstrate a comparison of MQB and qubit-only quantum resources, we apply our approach to pyrazine (\ce{C4N2H4}), a widely benchmarked system for nonadiabatic dynamics~\cite{seidner1992ab,woywod94,kuhl02}. 
In pyrazine, photoexcitation from the ground state to the $\pi\pi^*$ bright state decays to the $n\pi^*$ dark state through internal conversion, mediated by vibrational degrees of freedom dominated by a ring-breathing mode and an out-of-plane hydrogen wag~\cite{woywod94,kuhl02}. This nonadiabatic dynamics can be described with a linear-vibronic-coupling (LVC) Hamiltonian with two electronic states linearly coupled to two vibrational modes. Denoting the $\pi\pi^*$  state as $\ket{1}$ and $n\pi^*$ as $\ket{0}$, the pyrazine LVC Hamiltonian is
\begin{multline} \label{eq:ham-lvc}
	H_\mathrm{LVC} = \tfrac{1}{2} \sum_{j=1}^2 \omega_j ( P_j^2 + Q_j^2) - \tfrac{1}{2}\Delta E\, \sigma_z \\  + \kappa \sigma_z Q_1 + \lambda \sigma_x Q_2,
\end{multline}
where $\omega_1/(2\pi) = \qty{17.9}{\tera\hertz}$, $\omega_2/(2\pi) = \qty{28.5}{\tera\hertz}$, $\Delta E/(2\pi) = \qty{199}{\tera\hertz}$, $\kappa/(2\pi) = \qty{-30.7}{\tera\hertz}$, and $\lambda/(2\pi) = \qty{63.3}{\tera\hertz}$~\cite{kuhl02}.
The initial state $\rho_0$ is a Franck-Condon excitation from the ground state to the $\pi\pi^*$ state, corresponding to a displaced wavepacket with displacement parameter $\alpha = 0.210$~\cite{kuhl02}. To simulate pyrazine in an environment, we consider the case where the molecule is subjected to vibrational pure dephasing, i.e., $V_j = n_j$ in \cref{eq:lindblad-vc}, with rate $\gamma_j = \gamma_d = \SI{2.1e12}{\per\second}$ for each vibrational mode $j$~\cite{Stock_1990}. 

For MQB simulations, we model the simulator noise with a Lindblad master equation (\cref{eq:lindblad-vc}). For isolated pyrazine simulated on ion-trap hardware, native dissipation is not used as a resource and the dominant error source is vibrational pure dephasing~\cite{macdonell2023predicting,valahu2023direct,Olaya_2025}, which is modeled with Lindblad jump operators $V_j = n_j$ acting on each mode $j$ with the same rate $ \gamma_d^{\mathrm{nat}} = \gamma_d^{\mathrm{err}}$ (see \cref{fig:dynamics}a). The native noise rates depend on the hardware, but typical ion-trap values are $\gamma_d^{\mathrm{nat}} \in [\num{e0}, \num{e2}]\;\unit{\per\second}$~\cite{Brownnutt2015,macdonell2023predicting,valahu2023direct,Olaya_2025,lucas2007,Talukdar2016, Jarlaud2021}.

When simulating pyrazine in an environment, the vibrational pure dephasing of the simulator becomes usable dissipation. Based on recent MQB hardware~\cite{macdonell2023predicting,valahu2023direct,navickas2025experimental} we set $\gamma_d^{\mathrm{MQB}} = \gamma_d^{\mathrm{nat}} = \qty{30}{\per\second}$, giving $ F = \gamma_d^{\mathrm{nat}} / \gamma_d = \num{1.4e-11}$. Simulation errors are now due to the second-largest source of noise, which is not useful for describing our molecule-environment model. In ion-trap hardware, this is vibrational heating~\cite{macdonell2023predicting,valahu2023direct,Olaya_2025}, $V_j = a_j^{\dagger}$, with all modes subject to the same rate $\gamma_h^{\mathrm{nat}}=\gamma_h^{\mathrm{err}}$, typically in the range $\gamma_h^{\mathrm{nat}} \in [\num{e-1}, \num{e1}]\; \unit{\per\second}$~\cite{Brownnutt2015,macdonell2023predicting,valahu2023direct,Olaya_2025,lucas2007,Talukdar2016, Jarlaud2021}.

\Cref{fig:dynamics} shows the effect of the simulation errors on the accuracy of the two simulation approaches in reproducing the population dynamics $P_1(t)$ of the $\pi \pi^*$ state of isolated pyrazine. In the MQB simulation (\cref{fig:dynamics}a), decreasing the simulator noise $\gamma_d^{\mathrm{err}}$ decreases the error in $P_1(t)$, while in qubit-only simulation the error decreases with increasing Trotter number $N$. \Cref{appen:num_method} provides the details of our numerical implementation.

\begin{figure}
	\centering\includegraphics[width=\columnwidth]{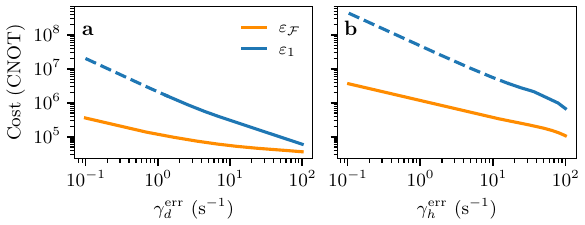} 
    \caption{MQB simulation requires orders of magnitude fewer quantum resources compared to qubit-only simulation with the same error. Shown is the equivalent computational cost, in logical CNOT gates, on a qubit-only computer for simulating the vibronic dynamics of~\textbf{(a)} isolated pyrazine and~\textbf{(b)} pyrazine in an environment (with $\gamma_d = \SI{2.1e12}{\per\second}$). The equivalent cost is defined by matching the error $\varepsilon_1$ or infidelity $\varepsilon_{\mathcal{F}}$ on the MQB simulator, which arise from pure vibrational dephasing $\gamma_d^{\mathrm{err}}$ in \textbf{(a)} and vibrational heating $\gamma_h^{\mathrm{err}}$ in~\textbf{(b)}. Typical ranges of these rates in trapped-ion simulators are $\gamma_d^{\mathrm{nat}} \in [\num{e0}, \num{e2}]$ \unit{\per\second} and $\gamma_h^{\mathrm{nat}} \in [\num{e-1}, \num{e1}]$~\unit{\per\second}~\cite{Brownnutt2015,macdonell2023predicting,valahu2023direct,Olaya_2025,lucas2007,Talukdar2016, Jarlaud2021}.
    Dashed lines are extrapolation.
    }
	\label{fig:cost}
\end{figure}

We estimate the cost of MQB simulations on trapped-ion hardware based on previous vibronic simulations~\cite{valahu2023direct,macdonell2023predicting,navickas2025experimental}. The memory cost is one qubit from a single ion to represent the two electronic states, as well as two bosons (two vibrations of the single ion) to encode the two molecular vibrational modes. The time cost is one gate because both the isolated and the in an environment simulations can be achieved with one gate, a single compound laser pulse~\cite{macdonell2023predicting}.

For qubit-only simulations, we determine the resources necessary to match the error of MQB simulators. The memory cost to simulate isolated pyrazine is 11 qubits: one to encode the two electronic states and five to encode each vibrational mode with Fock-space cutoff $d=32$. The time cost of each first-order Trotter step is 300 CNOT and 170 $R_z$ gates. Our qubit-only simulations of pyrazine in an environment consider the same case as our MQB simulations, namely vibrational pure dephasing in each of the two modes with $\gamma_d = \SI{2.1e12}{\per\second}$. Including the $J=2$ jump operators requires an additional $\lceil \log_2 (J+1)\rceil = 2$ ancilla qubits to encode $H_{\mathrm{dil}}$, giving a total memory cost of 13 qubits. The time cost per Trotter step is now 1800 CNOT and 700 $R_z$ gates.

To compare quantum resources between MQB and qubit-only simulators, we consider two error metrics that provide complementary information about simulation accuracy. The maximum infidelity between the approximate state $\tilde{\rho}(t)$ and the exact state $\rho(t)$ is
\begin{align}
    \varepsilon_{\mathcal{F}} & = \max_t \left( 1 - \mathcal{F}(t) \right) \\ 
    & =  \max_t \left( 1 - \left(\Tr \sqrt{ \sqrt{\rho(t)} \tilde{\rho}(t)\sqrt{\rho(t)}}\right)^2 \right),
\end{align}
and the maximum error in the $\pi\pi^*$ population is
\begin{equation}
    \varepsilon_1 = \max_t\big| \tilde{P}_1(t) - P_1(t)\big|,
\end{equation}
where $\tilde{P}_1(t)$ is the approximate and $P_1(t)$ the exact population. 
Infidelity is a global measure of how close the state remains to the exact one. By contrast, $P_1(t)$, like many chemical observables, probes local features of the state, making it less sensitive to simulation errors than the fidelity~\cite{Hauke_2012,Sarovar2017,heyl2019quantum}. Evaluating both metrics therefore gives a more complete picture of how MQB and qubit-only simulations perform.

We match the errors between MQB and qubit-only simulators to compare their costs fairly. In particular, we compute the qubit-only gate cost, in CNOT and $R_z$ gates, needed to achieve the same error as would be achieved with an MQB device with typical error rates. Specifically, given the $\gamma_d^{\mathrm{err}}$ rate of the MQB simulator, we determine the number of Trotter steps $N$ needed for the qubit-only simulator to achieve the same accuracy, as illustrated in \cref{fig:error_matching}. For pyrazine in an environment, we follow the same method, except that $\gamma_h^{\mathrm{err}}$ determines the MQB error. To match errors, we use cubic splines to interpolate between calculated values. To match low MQB simulation errors, we extrapolate the qubit-only results to larger Trotter numbers once numerical simulations become prohibitively expensive. The error of an observable is bounded by the operator-norm error for Trotter simulations, $\varepsilon_1 = O(t^2/N)$~\cite{Trotter,childs2021theory}, meaning that $\log(\varepsilon_1) = -\log N + \textrm{constant}$ at fixed $t$. We perform numerical simulations up to an $N$ large enough that $\varepsilon_1$ (and $\varepsilon_ {\mathcal{F}}$) follow this expected scaling, beyond which we use it for extrapolation, as shown in \cref{fig:error_matching}b.

Error-matching reveals the significant advantage of using bosons in the MQB approach, as shown in \cref{fig:cost}. For both error metrics, the cost of the qubit-only simulation with equivalent error increases as the MQB simulator error decreases. This result holds for both simulations of the isolated pyrazine molecule and in an environment.

\begin{table*}
\centering
\caption{Resource counts for simulating linear vibronic dynamics of pyrazine with two electronic states and two modes. MQB resources measure qudits and bosonic modes (memory) and sequential compound gates (time). Qubit-only resources measure qubits (memory) and CNOT and $R_z$ gates (time). Volume is memory multiplied by time (CNOT), and advantage is the ratio of qubit-only and MQB volumes. Results are for MQB simulator noise representative of recent MQB hardware~\cite{valahu2023direct,macdonell2023predicting}: $\gamma_d^{\mathrm{err}} = \qty{30}{\per\second}$ for the isolated molecule and $\gamma_h^{\mathrm{err}} = \qty{2e-1}{\per\second}$ for the one in an environment.} 
\label{tab:pyrazine-qubit-resources}
\vspace{3mm}
\begin{tabular}{llllll}
\toprule
 & MQB & \multicolumn{4}{c}{Qubit-only} \\
\cmidrule(lr){3-6}
 & & \multicolumn{2}{c}{Matched $\varepsilon_{\mathcal{F}}$} & \multicolumn{2}{c}{Matched $\varepsilon_1$} \\
\cmidrule(lr){3-4} \cmidrule(lr){5-6}
 & & Isolated & Open & Isolated & Open \\
\midrule
Memory & 3 & 11 & 13  & 11 & 13 \\
\midrule
Time & 1 & \num{4e4} CNOT & \num{3e6} CNOT & \num{1e5} CNOT & \num{2e8} CNOT \\ 
 & & \num{2e4} $R_z$ & \num{1e6} $R_z$ & \num{7e4} $R_z$ & \num{8e7} $R_z$ \\
\midrule
Volume & 3 & \num{4e5} & \num{3e7} & \num{1e6} & \num{2e9} \\ \midrule
MQB advantage & & \num{1e5} & \num{1e7} & \num{3e5} & \num{6e8} \\
\bottomrule
\end{tabular}
\end{table*}

\Cref{tab:pyrazine-qubit-resources} highlights specific costs for MQB noise levels reflective of recent experiments~\cite{macdonell2023predicting,valahu2023direct,navickas2025experimental}. For the isolated molecule, for qubit-only simulations to achieve the same infidelity as existing MQB devices requires at least \num{4e4} CNOT and \num{2e4} $R_z$ gates. Because the memory requirements are larger as well, the single-pulse MQB simulation equates to a QECV of \num{4e5}.

When environmental effects are included, the gap in quantum resources widens further by orders of magnitude. MQB simulations of a pyrazine molecule in an environment require far fewer resources than equally accurate qubit-only simulations. Specifically, we find that to match the infidelity of MQB simulations with one MQB gate requires \num{3e6} CNOT and \num{1e6} $R_z$ gates, giving a QECV of \num{3e7}. This improved MQB resource efficiency arises because bosonic degrees of freedom and their dissipators map naturally onto MQB simulators, while on qubit-only platforms these must be encoded into $H_{\mathrm{dil}}$ at significant cost.

The resource gap between MQB and qubit-only simulators is even larger for errors in the more chemically relevant population dynamics, $P_1(t)$. To match $\varepsilon_1$ in MQB simulations of isolated pyrazine, qubit-only simulations now require \num{1e5} CNOT and \num{7e4} $R_z$ gates, with a QECV of \num{1e6}. The resource gap between MQB and qubit-only simulations is enormous for pyrazine in an environment: $\num{2e8}$ CNOT and $\num{8e7}$ $R_z$ gates, with a QECV of over \num{2e9}. MQB simulators perform so well at calculating $P_1(t)$ primarily because dephasing in MQB devices damps coherences while preserving population transfer pathways, leading to small $\varepsilon_1$ even when global infidelity is modest. In contrast, Trotterization does worse because it induces coherent, history‑dependent distortions (phase shifts and amplitude bias) that directly perturb population dynamics, as shown in \cref{fig:dynamics}b. Therefore, for simulations of direct chemical relevance, such as populations, MQB devices can outperform qubit-only simulators by very large margins.

\section{Scaling with system size}
\label{sec:scaling}

Understanding how the resource requirements of MQB and qubit-only simulators compare with increasing system size is essential for assessing their performance on larger, classically intractable systems. We expect the resource advantage of MQB simulators over qubit-only ones to increase for medium-sized molecular problems, and likely more so for chemically relevant open-system cases, before ultimately being limited at very large sizes by the increasing errors in MQB devices.

We extend our resource comparison to larger molecules by adding modes to the LVC model of pyrazine, where each additional mode contributes one quadratic term $\omega_2(P_j^2 + Q_j^2)$ and one linear vibronic-coupling term $\lambda \sigma_x Q_j$. Because simulations beyond a few modes become prohibitively expensive on classical computers, we perform calculations at a smaller Fock-space truncation of $d=4$ for up to $M=5$ modes. MQB noise is modeled as a per-mode contribution, with each mode subject to the same dissipation rate---$\gamma_d$ for isolated molecules and $\gamma_h$ for molecules in an environment.

\begin{figure}
	\centering
    \includegraphics[width=\columnwidth]{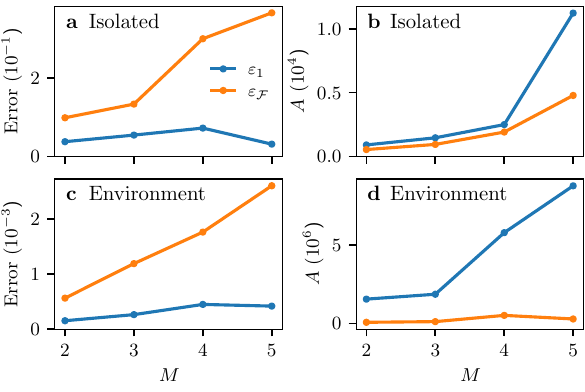} 
    \caption{The advantage of MQB simulators over qubit-only simulators grows with system size, even as MQB errors increase.
    \textbf{(a,c)}: MQB errors increase with the number of modes $M$ for both an isolated molecule and when in an environment (with $\gamma_d = \SI{2.1e12}{\per\second}$) because each mode experiences noise. Infidelity increases monotonically because it is a global error, whereas the population error is a local observable and can be non-monotonic.
    \textbf{(b,d)}: The advantage $A$---the ratio of qubit-only to MQB volume---increases with system size. Here, $\gamma_d^{\textrm{err}} = \SI{30}{\per\second}$ and $\gamma_h^{\textrm{err}} = \SI{0.2}{\per\second}$ for the isolated and in environment molecule, respectively.
    }
	\label{fig:scaling}
\end{figure}

MQB simulation further improves its advantage over qubit-only simulation with increasing system size, as shown in \cref{fig:scaling}. Although MQB simulation error grows with the number of modes (see \cref{fig:scaling}a,c), the resources required for qubit-only simulation to match the MQB error also increase, so long as the MQB error remains small, as in our simulations (see \cref{fig:scaling}b,d). Qubit-only resource costs rise both in qubit count (each additional mode introduces $5$ qubits) and in gate count (each mode contributes an additional sequence of CNOT ladders per Trotter step).

This advantage of MQB simulation at larger system sizes comes from the efficient encoding: the memory cost grows linearly as $1 + M$ because each additional mode requires only one additional boson, while the full time evolution can still be implemented with a single compound laser pulse, amounting to one gate~\cite{MacDonell2021}. However, the error also grows with system size: as more modes are included, each one experiences environmental noise intrinsic to the MQB device. If all modes experience comparable dissipation rates (as we assume), the total error is expected to scale at least linearly with system size, as is typical in other analog devices~\cite{Trivedi2024, Kashyap2025}.

Qubit-only devices encounter a different bottleneck, the rapid growth of gate cost with the number of modes. Although the qubit-only memory cost scales linearly with the number of modes as well, the time cost of the Trotter simulation to reach error $\varepsilon$ scales at least quadratically, as $O( M^{2+1/p} (\Lambda T)^{1+1/p} \varepsilon^{-1/p})$. This is because the gate cost grows linearly with $M$ and the Trotter number scales as $N = O( (M\Lambda T)^{1+1/p} \varepsilon^{-1/p})$. The latter is obtained from the commutator error for local Hamiltonians~\cite{childs2021theory}, where $H_{\textrm{qubits}}$ is $(d+1)$-local because the qubit register encoding the electronic states overlaps in support with each bosonic qubit register through the linear-vibronic-coupling terms. For local Hamiltonians and $p$th order product formulas, achieving error $\varepsilon$ requires $N = O(|||H_{\textrm{qubits}}|||_1 ||H_{\textrm{qubits}}||_1^{1/p} T^{1+1/p} \varepsilon^{-1/p})$ Trotter steps, where $|||\cdot|||_1$ denotes the induced 1-norm and $||\cdot||_1$ the 1-norm~\cite{childs2021theory}. Since $H_{\textrm{LVC}}$ has $\Theta(M)$ vibronic-coupling terms, $|||H_{\textrm{qubits}}|||_1 = O(M \Lambda)$ and $||H_{\textrm{qubits}}||_1 = O(M \Lambda)$, with $\Lambda = \max_k |\alpha_k|$.

For open systems, the resource advantage of MQB simulators is even greater as system size increases (see \cref{fig:scaling}d). Simulating more environmental interactions significantly raises resource costs in qubit-only devices; in particular, the gate count grows faster than for the isolated molecule because it scales with both the number of modes $M$ and the number of jump operators $J$. In contrast, although noise in MQB simulators is extensive, the same is true for molecules coupled to an environment. If the dominant noise sources are harnessed as part of the simulation, only the residual unwanted noise contributes to the error, which then grows more slowly than in simulations of isolated molecules and may remain sufficiently small, or otherwise be mitigated, to achieve useful accuracy.

\section{Discussion}

We have shown that MQB simulators require orders-of-magnitude less resources compared to qubit-only ones, even for small molecules such as pyrazine. The advantage becomes even larger when environmental effects are included and as system size increases. This resource efficiency arises because MQB simulators can natively represent electronic states (as a qudit), molecular vibrational modes (as bosons), and environmental effects (as Lindbladian dissipators).

Our resource counts are conservative, because all our algorithmic choices favor qubit-only efficiency. Most importantly, our qubit-only estimates only consider algorithmic error, i.e., they assume perfect, noiseless qubits and gates. In practice, incorporating fault tolerance would inflate both the qubit and gate counts by additional orders of magnitude~\cite{terhal2015quantum,litinski2019magic,Katabarwa2024}. Furthermore, our worst-case comparison is based on the infidelity $\varepsilon_{\mathcal{F}}$, which is particularly sensitive to MQB simulator noise; by contrast, errors in chemically relevant local observables, such as $\varepsilon_1$, would typically be significantly lower (see \cref{fig:cost}). Finally, we applied a generous $1/3$ reduction to CNOT counts to account for best-case circuit optimizations~\cite{Sawaya2020}, further biasing our estimates downward. For larger systems, qubit-only simulation approaches based on higher‑order product formulas~\cite{suzuki1990fractal,suzuki1991general,childs2021theory} or qubitization~\cite{Low2019,gilyen2019quantum,martyn2021grand} may offer improvements because they have better asymptotic scaling.

We expect the MQB advantage to persist for classically intractable system sizes. Under the analysis in \cref{sec:scaling}, the MQB resource advantage is expected to grow at least linearly with system size, since the resources required by qubit-only simulators scale at least quadratically, while errors in MQB devices increase approximately linearly. However, this growth cannot continue indefinitely. The analysis assumes small errors and therefore fails once the MQB error (whether $\varepsilon_{\mathcal{F}}$ or $\varepsilon_1$) becomes of order unity. No uncorrected simulator---quantum or classical---can simulate arbitrarily large systems for arbitrarily long times to an arbitrarily low error. However, the relevant question is whether practical, classically intractable problems can be solved in the foreseeable future given typical noise rates. We expect the answer to be yes because classical intractability arises in systems only slightly larger than those in our numerical simulations.

All results presented here are based on noise levels and capabilities representative of current MQB hardware. Further hardware improvements will extend the resource advantage to even larger molecular systems. Such improvements are likely to include improved experimental motional control and, in the longer term, bosonic quantum error suppression or correction~\cite{Weizhou2021, Brady2024} to reduce residual noise.

Overall, our results support a clear design principle for quantum simulation of chemistry: represent non‑qubit degrees of freedom natively whenever possible. For vibronic dynamics, this principle translates into orders‑of‑magnitude savings and a credible path to classically intractable regimes on near‑term MQB devices, especially for open‑system chemistry.

\begin{acknowledgments}
We thank Teerawat Chalermpusitarak, Mohammad Nobakht, Harsh Raj, Kai Schwennicke, Patrick Sinnott, Ting Rei Tan, and Christophe Valahu for valuable discussions.
We were supported by the Australian Research Council (FT230100653), by the Sydney Quantum Academy, by the United States Office of Naval Research Global (N62909-20-1-2047), and by Wellcome Leap.
\end{acknowledgments}

\section*{Appendices}

\appendix

\setcounter{section}{0}
\renewcommand{\thesection}{A\arabic{section}}%
\setcounter{equation}{0}
\renewcommand{\theequation}{A\arabic{equation}}%
\setcounter{figure}{0}
\renewcommand{\thefigure}{A\arabic{figure}}%
\setcounter{table}{0}
\renewcommand{\thetable}{A\arabic{table}}%

\begin{figure}[b]
	\centering
    \includegraphics[width=\columnwidth]{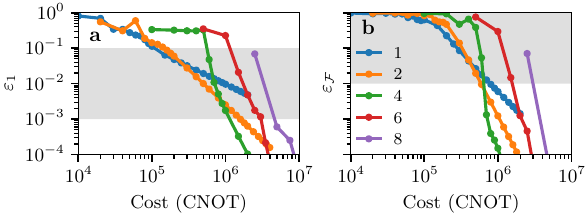} 
    \caption{Resource requirements for first- and second-order Trotterizations are comparable in qubit-only simulations when targeting \textbf{(a)} population error and \textbf{(b)} infidelity for isolated pyrazine at error levels typically achieved by MQB simulators (shaded region). The legend labels the $p$th-order Suzuki-Trotter decomposition.
    }
	\label{fig:trotter_order}
\end{figure}

\section{Higher-order Trotter}\label{appen:higher_trotter}

We evaluate qubit-only resource requirements to simulate the LVC model of pyrazine using first-order Trotterization, as it has resource costs comparable to higher-order schemes while remaining simpler to analyze, as shown in \cref{fig:trotter_order}. Although higher-order Trotter formulas offer better asymptotic error scaling, their large constant prefactors can make each step more costly. This creates a tradeoff where higher orders become advantageous only at very small target errors.

\section{Numerical implementation}\label{appen:num_method}

We simulated the MQB dynamics using the \verb|mesolve| function for open-system quantum dynamics in QuTiP~\cite{qutip5}. Our qubit-only simulations were obtained from a Trotter algorithm also implemented using QuTiP. \Cref{eq:ham-vc-qubit} was obtained numerically by using Cirq~\cite{cirq2025} to find equivalent Pauli sums and removing terms with coefficients $|\alpha_k| < \num{e-10}$, which occur due to numerical rounding errors. To estimate errors of our simulations, we compared the simulation results to numerically exact simulations obtained using the \verb|sesolve| and \verb|mesolve| functions in QuTiP for an isolated and open molecule, respectively. In all of our \verb|sesolve| and \verb|mesolve| simulations, we set the relative and absolute convergence tolerances to \num{e-8} and \num{e-6}, respectively.

\bibliography{bib}

\end{document}